\documentstyle[aps,twocolumn]{revtex}
\begin{document}
%\draft
\title{
Orientational Phase Transition in Solid $C_{60}$:
A Model Approach}
\author{T.I.Schelkacheva and E.E.Tareyeva}
\address{
Institute for High Pressure Physics, Russian Academy of Sciences, Troitsk
142092, Moscow region, Russia}
\date{\today}
\maketitle
\begin{abstract}
A simple model for the angular dependent interaction
between $C_{60}$ molecules in face centered cubic lattice is
proposed and analyzed by use of the rigorous bifurcation
approach. The quantitative results for the orientational phase
transition and the characteristics of the ordered phase are in
good agreement with the experimental data.

\end{abstract}


\narrowtext
%\twocolumn

The orientational phase transition in solid $C_{60}$ is of much
current interest. The orientational ordering has been a subject
of extensive experimental investigations (see, e.g.,
~\cite{dav,hein,dav1,blinc,blas,braj} ); some theoretical
researches were performed, too
~\cite{mich,yha,lu,heid,lap,laun,gam}.  However, {\it ab initio}
calculations fail to reproduce the experimental results.

In this Letter we develop a simple model for the angular
dependence of the intermolecular potential in solid $C_{60}$. The
model is based on the ideas of preferred orientations due
to David et al. ~\cite{dav} and to Lapinskas et al. ~\cite{lap},
and on the maximal exploit of symmetry considerations.
We apply to this model interaction the rigorous approach
based on the Lyapunov--Schmidt theory of bifurcation of
solutions of nonlinear integral equations and obtain
quantitative results for the transition temperature and
the distribution of
molecular orientations in the ordered phase. These results
occur to reproduce the experimental data.

As is established in a number of experiments $C_{60}$
crystallizes in a face centered cubic (fcc) structure.
At ambient temperature the
molecules rotate almost freely with centers on the fcc lattice
sites, so that the space group is $Fm{\bar 3}m$ (see, e.g.,
~\cite{hein}). When the temperature decreases to
$T_S \approx 260 K$ the first order orientational phase
transition takes place: the sites of the initial fcc 
latice become divided between four simple cubic sublattices
(see fig.1) 
\begin{figure}
%
%\vspace{-1.5cm}
%
\begin{picture}(180,140)(0,20)
\put(90,95){\vector(0,2){30}}
\put(90,95){\vector(2,0){30}}
\put(90,95){\vector(-1,-1){22}}
\put(90,95){\circle*{20}}
\put(90,135){$z$}
\put(125,95){$y$}
\put(65,65){$x$}
\put(100,100){0}
\put(60,60){\line(0,2){37}}
\put(60,113){\line(0,2){37}}
\put(60,105){\circle{16}}
\put(57,103){A}
\put(30,30){\line(0,2){37}}
\put(30,83){\line(0,2){37}}
\put(27,73){A}
\put(30,75){\circle{16}}
\put(150,60){\line(0,2){37}}
\put(150,113){\line(0,2){37}}
\put(150,105){\circle{16}}
\put(147,103){A}
\put(120,30){\line(0,2){37}}
\put(120,83){\line(0,2){37}}
\put(120,75){\circle{16}}
\put(117,73){A}

\put(60,60){\line(2,0){37}}
\put(113,60){\line(2,0){37}}
\put(105,60){\circle{16}}
\put(102,58){B}
\put(30,30){\line(2,0){37}}
\put(83,30){\line(2,0){37}}
\put(75,30){\circle{16}}
\put(72,28){B}
\put(30,120){\line(2,0){37}}
\put(83,120){\line(2,0){37}}
\put(75,120){\circle{16}}
\put(72,118){B}
\put(60,150){\line(2,0){37}}
\put(113,150){\line(2,0){37}}
\put(105,150){\circle{16}}
\put(102,148){B}

\put(150,60){\line(-1,-1){10}}
\put(130,40){\line(-1,-1){10}}
\put(135,45){\circle{16}}
\put(132,43){D}
\put(60,60){\line(-1,-1){10}}
\put(40,40){\line(-1,-1){10}}
\put(45,45){\circle{16}}
\put(42,43){D}

\put(60,150){\line(-1,-1){10}}
\put(40,130){\line(-1,-1){10}}
\put(45,135){\circle{16}}
\put(42,133){D}

\put(150,150){\line(-1,-1){10}}
\put(130,130){\line(-1,-1){10}}
\put(135,135){\circle{16}}
\put(132,133){D}
\put(70,5){Fig.1.}
\end{picture}
\end{figure}
\noindent
with own preferable molecular orientation in each
sublattice. The broken symmetry space group is $Pa{\bar 3}$.

Moreover, the neutron--diffraction experiments
~\cite{dav} have shown that the orientations in the
ordered state are so that the electron--rich regions (the
interpentagon double bonds) face the electron--deficient regions
of the neighboring $C_{60}$ molecule: the centers of pentagons
or the centers of hexagons.  It was shown
~\cite{dav,dav1,blas} that the
ratio of the number of molecules in those two states is about
$60:40$ at the phase transition temperature and increases when
the temperature decreases. This remaining orientational disorder
is usually believed to cause the orientational glass transition
at $T_G \approx 90K$ now confirmed by various experimental
technics (see, e.g., ~\cite{glass}).  These two minima of the
intermolecular angle dependent energy were obtained by numerical
calculations and were shown to be much lower than the energies
of other mutual orientations of the pair of molecules (see,
e.g., ~\cite{yha,lu,gam,yha1}). In those calculations the
previously obtained charge distribution for the isolated
$C_{60}$ molecule ~\cite{mol} was taken into account.
 Usually recent calculations use the intermolecular
potential of Sprik et al. ~\cite{sprik}: a sum
of 6-12 and Coulomb interactions between 60 atoms $C$
and 30 double--bond centers $D$ and between each other:
\begin{eqnarray}
\Phi(1,2) & = &
\sum_{k\in C(1)} \sum_{k'\in C(2)}
4 \epsilon \left \{ \left( \frac {\sigma_{CC}} {R_{kk'}}
\right)^{12} - \left( \frac {\sigma_{CC}}
{R_{kk'}}\right)^6\right\} \nonumber\\
& + &\sum_{k \ne k', k, k' \in C,D}
4 \epsilon \left \{ \left( \frac {\sigma_{CD}} {R_{kk'}}
\right)^{12} - \left( \frac {\sigma_{CD}}
{R_{kk'}}\right)^6\right\} \nonumber\\
& + &\sum_{k\in D(1)} \sum_{k'\in D(2)}
4 \epsilon \left \{ \left( \frac {\sigma_{DD}} {R_{kk'}}
\right)^{12} - \left( \frac {\sigma_{DD}}
{R_{kk'}}\right)^6\right\} \nonumber\\
& + & \sum_{k,k'\in C,D} \frac {q_{k} q_{k'}} {R_{kk'}}
\label{full}
\end{eqnarray}
Here
$\epsilon = 1.293 meV, \sigma_{CC} = 3.4 \AA, \sigma_{CD} = 3.5
\AA, \sigma_{DD} = 3.6 \AA, q_D = - 0.35e, q_C = - q_D/2.$

Rigorously speaking we are interested in the angular part of
this complicated interaction represented in terms of
multipole--multipole interaction of point--like multipoles on the
sites of rigid fcc lattice with coefficients to be calculated
from ~(\ref{full}).
The general form of this angular part is
\begin{equation}
\Phi_{ij} (\omega_i, \omega_j) =
\sum_{l;\nu, \tau}
C_{\nu, \tau}^l (\omega_{ij})  u _{l \nu} (\omega_i) u_{l \tau}
(\omega_j),
\label{mult}
\end{equation}
with
$l = 6, 10, 12, 16, 18, ...$
due to the icosahedral molecular symmetry
$I_h$.
In the Eq.~(\ref{mult})   $\omega_i$ are the angles describing
the orientation of the molecule on site $i$, for example, Euler
angles and $u_{l \nu }$ -- some kind of harmonics.  However, we
simplify the problem and develop a model orientational
interaction.  As the angular dependent interaction is rather
short--ranged we can restrict ourselves by the nearest--neighbor
interactions.

We follow the main ideas of the papers ~\cite{dav,lap}
and use the restricted number of allowed orientations instead
of free continuous rotations.
Let us take into account in the energy ~(\ref{mult}) only
the orientations
with pentagons, hexagons or double bonds directed
towards 12 nearest neighbors in fcc lattice.
The $C_{60}$ molecule is constructed in such a way that if
6 of its 12 pentagons (or 6 of its 20 hexagons) face 6
nearest neighbors double bonds ($P$ and $H$ states of Lapinskas
et al.~\cite{lap}),  then 6 of its 30 interpentagon double bonds
face the remaining 6 nearest neighbors. Now the
energy matrix elements
can take only three values: $J_0$ -- the
energy of the general mutual position, $J_P$ -- pentagon
{\it versus} double bond and $J_H$ -- hexagon {\it versus}
double bond. These energies in our model
can be compared with those calculated in
~\cite{yha,lu,gam,yha1}
as functions of the angular displacements
of the molecule at (0,0,0). Following ~\cite{lap}, and putting
$J_0 = 0$ we obtain from the fig.2(b) of the paper ~\cite{yha}
$J_P = - 300K$ and $J_H = - 110K$. Now we leave the
paper ~\cite{lap} and follow our own way.

The energy matrix elements $J_P$ and $J_H$ connect the
states of molecules only in the allowed orientations.
So, only allowed linear combinations of
$u_{l \nu}$ enter the Eq. ~(\ref{mult}).
The theoretical curve in ~\cite{yha} makes no difference
between the number $l$ of harmonics and describes
the effect of all of them. So,
in the framework of our model calculation it is possible
to build up the allowed functions using only the harmonics
with $l=6$: we need only their transformation properties. We
restrict ourselves to $l=6$, however the coefficients $J_P$ and
$J_H$ are not some of $C_{\nu ,\tau }^6$ given in ~(\ref{mult})
but effectively take into account higher order terms.
Let us construct the functions $P_i(\omega )$ and $H_i(\omega )$
explicitly in terms of cubic harmonics
$K_{m} \equiv K_{6,m}, m = 1, 2, ..., 13$
(see, e.g., ~\cite{heid}). All functions
$P_i$ and $H_i$ are the sums of $K_m$, invariant under
the icosahedral
symmetry of the molecule (i.e. belonging to the $A_{1g}$
representation of the icosahedral group $I_h$) if 
icosahedrons are naturally oriented in one
of 8 properly chosen coordinate systems.
The states $P_i (H_i)$ have 6 pentagons (hexagons)
and 6 double bonds directed towards 12 nearest
neighbors along different $[100]$ axes. $P_1(\omega )$
describes the molecule rotated from the standard
orientation $B$ (following ~\cite{harr}) about $[111]$
axis through the angle $97.76125^o$.
The angle for $H_1(\omega )$ is $37.76125^o$.
The functions $P_2 (\omega
)$,$P_3 (\omega )$ and $P_4 (\omega )$ (or $H_2,
H_3, H_4$) are obtained from $P_1(\omega )$ $(H_1)$ by
subsequent counter--clockwise rotations of the molecule by
$90^o$ around $z$ axes.

If written in the standard coordinate frame with
Cartesian axes along the cube sides these functions have the
following explicit form:
\begin{eqnarray}
P_{1}(\omega ) & = &
\alpha_{P} K_{1} (\omega ) + \beta_{P} [K_{8}(\omega ) + K_{9}
(\omega ) + K_{10} (\omega )] \nonumber \\ & + & \gamma_{P}
[K_{11}(\omega ) + K_{12} (\omega ) + K_{13} (\omega )],\nonumber\\
P_{2}(\omega ) & = & \alpha_{P} K_{1} (\omega ) +
\beta_{P} [ - K_{8}(\omega ) + K_{9} (\omega ) - K_{10} (\omega
)] \nonumber\\& + & \gamma_{P} [ - K_{11}(\omega ) + K_{12}
(\omega ) - K_{13} (\omega )], \nonumber\\ P_{3}(\omega ) & = &
\alpha_{P} K_{1} (\omega ) + \beta _{P} [K_{8}(\omega ) - K_{9}
(\omega ) - K_{10} (\omega )] \nonumber\\& + & \gamma _{P}
[K_{11}(\omega ) - K_{12} (\omega ) - K_{13} (\omega )],
\nonumber\\
P_{4}(\omega ) & = & \alpha_{P} K_{1} (\omega ) + \beta _{P}
[ - K_{8}(\omega ) - K_{9} (\omega ) + K_{10} (\omega )]
\nonumber\\& + &
\gamma _{P}
[ - K_{11}(\omega ) - K_{12} (\omega ) + K_{13} (\omega )],
\label{ph}
\end{eqnarray}
with $\alpha_{P}=-0.38866;$ $\beta _{P}=0.31486;$
$\gamma _{P}=-0.42877.$ The functions $H_i(\omega )$
have the same form as $P_i(\omega )$ but with the coefficients
$\alpha _{H}= 0.46588;$ $\beta _{H}=0.37740;$
$\gamma _{H}=0.34432$. The functions are normalized to unity.

Let us now treat our model by use of bifurcation approach
in the mean--field approximation.
As is well known the mean--field approach often brings one
to the formulation of the broken space symmetry problem
in terms of the bifurcation of solutions of nonlinear
integral equations for distribution functions (see, e.g., the
review ~\cite{koz}). In particular, the bifurcation approach was
used in the case of orientational phase transitions in molecular
crystals in the Refs.\onlinecite{tt1,ttdan,tt2,rlem,shsh,sch}
etc.  The simplified version was originally developed by James
and Keenan for solid methane ~\cite{jkee} and by Michel, Copley
and Neumann ~\cite{mich} and by Heid ~\cite{heid} for solid
$C_{60}$. We shall follow our papers on hydrogen
~\cite{tt1,ttdan,tt2,sch}.

In the mean--field approximation from the first
equation of BBGKY hierarchy for the orientational distribution
functions or by minimizing the orientational free energy
one can obtain the following nonlinear integral equation
\cite{ttdan}:
\begin{equation}
g_i ( \omega_i) + \frac{1} {\Theta} \sum_{i \ne j} G_j
\int d \omega_j \Phi_{ij} ( \omega_i, \omega_j)
 e^{g_i (\omega_j)} = 0;
\label{main}
\end{equation}
$g_i (\omega_i) = \ln[\frac {f_i (\omega_i)} {G_i}]$,
$f_i ( \omega_i)$ -- one--particle orientational distribution
function for a molecule on $i$-th lattice site, the constants $
G_i$ are the normalization constants.

In our case of solid $C_{60}$ where there are four sublattices
(see fig.1) and four kinds of unknown distribution functions
we obtain from ~(\ref{main}) the following system of four
nonlinear integral equations:

\begin{eqnarray}
g_1(\omega )& + &\lambda \int d \omega ' [
B(\omega ,\omega ') G_2 e^{g_2(\omega ')} + A(\omega ,\omega ')
G_3 e^{g_3 (\omega ')}
\nonumber\\& + &
D(\omega ,\omega ') G_4 e^{g_4 (\omega
')}] = 0,
\nonumber\\
g_2(\omega )& + &\lambda
\int d \omega ' [ B(\omega ,\omega ') G_1 e^{ g_1(\omega ')} +
A(\omega ,\omega ') G_4 e^{g_4 (\omega ')}
\nonumber\\& + &
D(\omega ,\omega ')
G_3 e^{g_3 (\omega ')}] = 0,\nonumber\\
g_3(\omega )& + &\lambda \int d \omega ' [ B(\omega ,\omega ')
G_4 e^{g_4(\omega ')} + A(\omega ,\omega ') G_1 e^{g_1 (\omega
')}
\nonumber\\& + &
D(\omega ,\omega ') G_2 e^{g_2 (\omega ')}] = 0,
\nonumber\\
g_4(\omega )& + &\lambda \int d \omega'
[ B(\omega ,\omega ') G_3 e^{g_3(\omega ')} + A(\omega ,\omega ')
G_2 e^{g_2 (\omega ')}
\nonumber\\& + &
D(\omega ,\omega ') G_1 e^{g_1 (\omega
')}] = 0.
 \label{int}
\end{eqnarray}
Here $\lambda=1/T$, 
$A(\omega,\omega')$, $B(\omega ,\omega ')$, $D(\omega ,\omega ')$
are the sums of interactions over
nearest neighbors in the sublattices $A$, $B$ and $D$ (see
fig.1), respectively.  For example, the sum in the plain
perpendicular to the $x$ axis can be written explicitly
in the form:
\begin{eqnarray}
D(\omega ,\omega ') & = & \nonumber\\
=2 \{[(P_1 (\omega ) &+& P_4 (\omega )]
J_{P} + (H_1 (\omega ) + H_4 (\omega _)] J_H]\nonumber\\
 \times  [
P_2 (\omega ')& +& P_3 (\omega ') + H_2 (\omega ') + H_3 (\omega
')]\nonumber\\
+  [P_2 (\omega ) &+& P_3 (\omega ) + H_2
(\omega ) + H_3(\omega )] \nonumber\\
\times  [(P_1 (\omega ') &+&
P_4 (\omega ')) J_P + (H_1 (\omega ') + H_4 (\omega ')) J_H]
\nonumber\\
 +  [(P_2 (\omega ) &+& P_3 (\omega )) J_P + (H_2
(\omega ) + H_3 (\omega )) J_H]\nonumber\\
\times  [ P_1 (\omega ') &+& P_4 (\omega ') + H_1 (\omega ') +
H_4 (\omega ')] \nonumber\\
 +  [P_1(\omega ) &+& P_4 (\omega )
+ H_1 (\omega ) + H_4 (\omega )]\nonumber\\
 \times
[(P_2(\omega ') &+& P_3(\omega ')) J_P + (H_2 (\omega ') + H_3
(\omega ')) J_H]\}
\label{pph}
\end{eqnarray} and analogously
for two other sublattices.

The equations ~(\ref{int}) are well known
Hammerstein equations ~\cite{ham}. In the case of finite
domain of integration  when the fixed point principle is valid
there exists detailed theory for such equations
(see ~\cite{kra}). We use the standard methods
(see, e.g. ~\cite{tren}). At high temperature the system
~(\ref{int})
has only trivial solution
 $g_i( \omega_i)=0$, corresponding to the
orientationally disordered phase. At the bifurcation points
$\lambda _{\alpha}$  new solutions with broken symmetry
appear ( $\lambda _{\alpha} > 0$).
For $\lambda = \lambda _ {\alpha} (1+ \mu)$ the functions
$g_i^{\alpha }( \omega_i)$ can be written as series in
integer or fractional powers of
$\mu$. These powers are defined by the bifurcation
equation (see ~\cite{tren}) corresponding to the system
~(\ref{int}).
In our case we have
$$g_i (\omega ) = \mu h_i (\omega ) + \mu^2 x_i (\omega ) +
...$$
 because among the integrals $\int d \omega K_{m_1} K_{m_2}
K_{m_3}$ with $m=8,...,13$ there are some which are not equal
to zero. This means the first order phase transition
~\cite{tt1,ttdan,tt2}.

The bifurcation points are the eigenvalues $\lambda _ \alpha$ of
the linearized system corresponding to
~(\ref{int}):
\begin{eqnarray}
h_1(\omega ) &+& \frac {\lambda} {4 \pi} \int d \omega ' [
B(\omega ,\omega ') h_2(\omega ') + A(\omega ,\omega ') h_3
(\omega ') \nonumber\\
& + &D(\omega ,\omega ') h_4 (\omega ')] = 0.\nonumber\\
\dots &\dots & \hbox to 6cm{\dotfill}
\label{sys}
\end{eqnarray}
The functions $h_i$ can be written in the form
\begin{equation} h_i(\omega) = \sum_{\nu} h_{i}^{\nu} K_{\nu}
(\omega), \label{hk} \end{equation} so that the eigenvalues
$\lambda _ \alpha $ define not only the bifurcation temperatures
but the relations between nonzero coefficients $h_{i}^{\nu}$
(that is the symmetry of the new phase), too.

In the case of the full interaction one can obtain all
possible broken symmetry phases compatible with
the initial symmetry and the condition of positive
temperature value (see, e.g., the case of hydrogen
~\cite{tt1,tt2,sch}). Now we have truncated the interaction
and reduced the problem. Nevertheless, there remain still
two quantitative characteristics we aim to obtain:
the bifurcation temperature and the relation between the
weights of $P$ and $H$ functions in the solution.

Using  ~(\ref{hk}) it is easy to rewrite the system ~(\ref{sys})
as the system of linear algebraic equations for the coefficients
$h_i^{\nu}$ with $i=1,2,3,4$ and $\nu = 1,8,9,...,13$.

Using the explicit form of the matrices
$A, B, D$  it is easy to obtain the only nonzero elements:
$$A_{1,1}  =  B_{1,1} = D_{1,1} \equiv u,$$
$$A_{8,8}  =  B_{9,9} = D_{10,10} \equiv v,$$
$$A_{11,11}  =  B_{12,12} = D_{13,13} \equiv z,$$
$$A_{8,11} =  A_{11,8} = B_{9,12} = B_{12,9} = D_{10,13} =
D_{13,10} \equiv w.$$
One can write the elements $u, v, z$ and $w$
in terms of the coefficients $\alpha _P,  \beta
_P, \gamma _P, \alpha _H, \beta _H, \gamma _H$ and energies
$J_P, J_H$ and obtain the following values: $u = 32 \cdot
5.046, v = 32 \cdot 94.127, z = 32 \cdot 7.665, w = - 32
\cdot 37.155$.

The determinant of the algebraic system is factorized in
$2 \times 2$  determinants, so that the eigenvalues
$\lambda _ \alpha $ can be easily obtained. Among the values
$\lambda _ \alpha $ there are two positive values.
The first one $\lambda_1  = 4 \pi /u$ corresponds to the solution
proportional to $K_1$ and is of no interest now.
The second is the positive solution of the equation
\begin{equation}
1 - \frac {\lambda} {4 \pi} (v + z) + \frac
{{\lambda}^2} {(4 \pi)^2} (vz - w^2) = 0,
\label{eql}
\end{equation}
namely
$ \lambda_b = 0.00364 K^{-1}$ or $ T_b = 275 K$.
The corresponding nontrivial eigenfunctions have
$h_i^1 = 0,$ and the other coefficients $h_i^{\mu }$
are subject to some constraints.
If we add
the condition for the functions $h_i(\omega )$ to transform one
into another under the action of the cubic group rotation
elements which leave the fcc lattice invariant, then
only three of the coefficients remain to be independent and the
functions $h_i$ can be written in the following form:
\begin{eqnarray}
h_1(\omega ) & = & a P_1 (\omega ) + b H_1 (\omega ) + c K_1
(\omega),\nonumber\\
h_2(\omega ) & = & a P_3 (\omega ) + b H_3 (\omega ) + c K_1
(\omega),\nonumber\\
h_3(\omega ) & = & a P_4 (\omega ) + b H_4 (\omega ) + c K_1
(\omega),\nonumber\\
h_4(\omega ) & = & a P_2 (\omega ) + b H_2 (\omega ) + c K_1
(\omega),
\label{hpk}
\end{eqnarray}
\begin{equation}
a \alpha _P + b \alpha _H + c = 0,
\label{fin1}
\end{equation}
\begin{equation}
a \beta _P + b \beta _H = Q (a \gamma _P + b \gamma _H),
\label{fin2}
\end{equation}
$$ Q = \frac {1 - v \frac
{\lambda} {4 \pi}} { \frac {\lambda} {4 \pi}w}.$$

Using the numerical value for $Q$
we obtain immediately:
\begin{equation}
\rho _P =\frac {a} {a + b} = 0.608; \qquad \rho _H=\frac {b} {a
+b} = 0.392,
\label{ro}
\end{equation}
so that not only the  transition temperature, but also the ratio
of the number of molecules in $P$ and $H$ states occur to
coincide with the experimental data ~\cite{dav,dav1,blas}
$\rho _P:\rho _H=60:40.$

To obtain the remaining unknown coefficient we use
the equations of the second order in $\mu$
($ g_i (\omega ) = \mu h_i (\omega ) + \mu ^2 x_i (\omega );
\lambda  = \lambda _c (1 + \mu )$).
The system has the form: $ \hat L \times {\bf x} = {\bf R} $,
where $\hat L$ is the linear $4 \times 4$ operator
 $\hat L \times {\bf h} = 0.$  The system has
nontrivial solutions for $x_i$ if the right hand side is
orthogonal to the solutions for $h_i$ obtained before. All 16
equations
\begin{equation}
\int d\omega R_i (\omega ) h_j (\omega )
= 0
\label{fin3}
\end{equation}
are identical due to the symmetry of
coefficients.
Solving equations
~(\ref{fin1}), ~(\ref{fin2}) and
~(\ref{fin3}), we obtain finally
$ a = - 25.7;\quad b =  - 16.6;\quad c = -2.27.$
The minus sign means that the solution goes
in the direction of higher temperatures ($ \mu =
- \tau, \tau = (T - T_b)/T_b$). The solution has
the turning point $T_t$ which is some Kelvins higher than
$T_b$.  The actual first order phase transition obtained from
the free energy behavior takes place between these two points.
The details of this calculation are to be published elsewhere.

To conclude, we developed a simple model for angle
dependent interaction for $C_{60}$ molecules in
the fcc cubic lattice. We used rigorous analytic approach
based on the Lyapunov--Schmidt theory of bifurcation
of solutions of nonlinear integral equations
to treat this model. As result we obtained the first
order phase transition, the bifurcation temperature
$ T_b = 275 K$, the $Pa{\bar 3}$  symmetry of the ordered
phase and the
ratio $\rho $ of the number of molecules with pentagon facing
neighbor double bond to the number of molecules with hexagon
facing neighbor double bond ~(\ref{ro}) near the phase
transition in good agreement with the experimental data.

This work was partially supported by Russian Foundation for
Basic Researches (Grant No. 98-02-16805).

The authors would like to thank V.N.Ryzhov and V.A.Davydov for
useful discussions.


\begin{references}
\bibitem{dav} W.I.F.David, R.M.Ibberson, T.J.S.Dennis, J.P.Hare,
and K.Prassides, Europhys. Lett., {\bf 18}, 219 (1992).
\bibitem{hein} P.A.Heiney,G.B.M.Vaughan,J.E.Fischer,N.Coustel,\\
D.E.Cox,J.R.D.Copley,D.A.Neumann, \\
W.A.Kamitakahara,K.M.Creegan,D.M.Cox,\\J.P.McCauley,
Jr.,A.B.Smith III, Phys.Rev., {\bf B 45}, 4544 (1992).

\bibitem{dav1} W.I.F.David, Europhys.News, {\bf 24}, 71 (1993).
\bibitem{blinc} R.Blinc, J.Selinger, J.Dolinsek, and D.Arcon,
Phys.Rev., {\bf B 49}, 4993 (1994).

\bibitem{blas} O.Blaschk, G.Krexner, Ch.Maier,R.Karawatzky,\\
Phys.Rev. {\bf B 56}, 2288 (1997).

\bibitem{braj} V.V.Brazh kin, and A.G.Liapin, UFN (Uspechi, 
Moscow) {\bf 166}, 893 (1996).  
\bibitem{mich} K.H.Michel,
J.R.D.Copley, and D.A.Neumann,\\ Phys.Rev.Lett. {\bf 68}, 2929 
(1992).  
\bibitem{yha} T.Yildirim and A.B.Harris, Phys.Rev. {\bf 
B 46}, 7878 (1992).  
\bibitem{lu} J.P.Lu, X.-P.Li, and 
R.M.Martin, Phys.Rev.Lett., {\bf 68}, 1551 (1992).  
\bibitem{heid} R.Heid, Phys.Rev. {\bf B 47}, 15912
(1993).
\bibitem{lap} S.Lapinskas, E.E.Tornau, and A.Rosengren,
Phys.Rev.{\bf B 49}, 9372 (1994).
\bibitem{laun} P.Launois,
S.Ravy, and R.Moret, Phys.Rev.{\bf B 55}, 2651 (1997).
\bibitem{gam} Z.Gamba, Phys.Rev. {\bf B 57}, 1402 (1998).
\bibitem{glass}
F.Gugenberger,R.Heid,C.Meingast,P.Adelman,M.Braun,\\
H.W\"uhl,M.Haluska,and
H.Kuzmany, Phys.Rev.Lett. {\bf 69}, 3774 (1992).
\bibitem{yha1}
T.Yildrim, A.B.Harris, S.C.Erwin, and M.R.Pederson,\\
 Phys.Rev.
{\bf B 48}, 1888 (1993).
\bibitem{mol} B.P.Feuston, W.Andreoni, M.Parinello, \\
and E.Clementi, Phys.Rev. {\bf B 44}, 4056 (1991);\\ N.Troullier
and J.L.Martins, Phys.Rev. {\bf B 46}, 1754 (1992), and
references therein.
\bibitem{sprik} M.Sprik, A.Cheng, and
M.L.Klein, J.Phys.Chem.  {\bf 96}, 2027 (1992).
\bibitem{harr} A.B.Harris, Physica {\bf A 205}, 154 (1994).
\bibitem{koz}J.J.Kozak, Adv.Chem.Phys., {\bf 40}, 229 (1979).
\bibitem{tt1}
E.E.Tareyeva, and T.I.Trapezina, Physics  Letters, {\bf A51},
114 (1975); {\it ibid} , {\bf A60}, 217 (1977).
\bibitem{ttdan}
E.E.Tareyeva, and T.I.Trapezina, DAN USSR (Soviet Phys.
Doklady), {\bf 223}, 823 (1975); Theor.Math.Phys. (Moscow), {\bf
26}, 269 (1976).
\bibitem{tt2} E.E.Tareyeva and
T.I.Schelkacheva, Theor.Math.Phys. (Moscow), {\bf 31}, 359
(1977); V.N.Ryzhov, E.E.Tareyeva and T.I.Schelkacheva, {\it ibid
} {\bf 40}, 269 (1979).
\bibitem{rlem} H.L.Lemberg, and S.A.Rice,
Physica, {\bf 63}, 48 (1973).
\bibitem{shsh} A.N.Shramkov, and P.P.Shtifanuk,
Theor.Math.Phys. (Moscow), {\bf 90}, 246 (1992).
\bibitem{sch}
T.I.Shchelkacheva, Physics Letters, {\bf A239}, 397 (1998).
\bibitem{jkee} H.M.James,and T.A.Keenan,J.Chem.Phys.{\bf 31},12(1959).
\bibitem{ham} A.Hammerstein, Acta Math., {\bf 54}, 117 (1930).
\bibitem{kra} M.A.Krasnosel'skii, Topological methods in the
theory of Nonlinear Integral Equations, Pergammon press, London,
1964.
\bibitem{tren} M.Vainberg and V.A.Trenogin, Russian Math.
Surveys, {\bf 17}, 1 (1962); M.Vainberg, {\it Variational
Methods for the Study of Nonlinear Operators}, Holden Day,
San Francisco, 1964.
\end{references}
\end{document}